\documentclass[twocolumn,showpacs,preprintnumbers,amsmath,amssymb]{revtex4}
\usepackage{dcolumn}
\usepackage{bm}
\usepackage{amsmath}
\usepackage[dvipdf]{graphicx}

\usepackage{epsfig,color}
\usepackage{amsfonts,color}

\definecolor{brown}         {rgb}{0.65 , 0.16 , 0.16}
\definecolor{orange}        {rgb}{1.00 , 0.60 , 0.00}
\definecolor{violet}          {rgb}{0, 0 , 0.5}

\DeclareGraphicsExtensions{.jpg,.pdf,.mps,.png,.eps,.ps,.EPS}
			   
\begin{document}
\def\be{\begin{equation}}
\def\ee{\end{equation}}
\def\bc{\begin{center}}
\def\ec{\end{center}}
\def\bea{\begin{eqnarray}}
\def\eea{\end{eqnarray}}
\def\mx{{\mathbf x}}

\title{A topological approach to neural complexity} 
\author{M. De Lucia$^{1,3}$, M. Bottaccio$^{1,3}$,
M. Montuori$^{1,3}$, L.Pietronero$^{1,2,3}$} \affiliation{$^1$INFM
SMC-Dipartimento di Fisica Universit\`a ``La Sapienza'', P.le A. Moro
5, 00185 Roma, Italy} \affiliation{$^2$ Dipartimento di Fisica
Universit\`a ``La Sapienza'', P.le A. Moro 5, 00185 Roma, Italy}
\affiliation{$^3$ Centro Fermi,Compendio Viminale, Roma, Italy }

\begin{abstract}
Considerable efforts in modern statistical physics is devoted to the
study of networked systems. One of the most important example of them
is the brain, which creates and continuously develops complex networks
of correlated dynamics. An important quantity which captures
fundamental aspects of brain network organization is the neural
complexity $C(\mathbf{X})$ introduced by Tononi {\it et al.}.  This
work addresses the dependence of this measure on the topological
features of a network in the case of gaussian stationary process. Both
analytical and numerical results show that the degree of complexity
has a clear and simple meaning from a topological point of
view. Moreover the analytical result offers a straightforward and
faster algorithm to compute the complexity of a graph than the
standard one.
\end{abstract}
\pacs{: 89.75.Hc, 87.80.Tq, 89.75.Fb}
\maketitle

I. INTRODUCTION\\

The study of networked systems such as the Internet, social networks
and biological networks has recently attracted great interest within
the statistical physics community. A large variety of techniques and
models have been developed in order to understand or predict the
behavior of these systems.  Great efforts have been applied in
discovering their topological features \cite{ref1,ref2,ref3,ref4} and
how these properties influence the behaviour of dynamical processes
taking place on them.  For example, we would like to know how the
topology of social networks influences the spread of information
\cite{social1,social2}, how the search engines are affected by World
Wide Web structure \cite{www1,www2}.\\ In this paper we focus
on a first basic approach for studying the interplay between dynamics
and topology of brain networks.This study has great interest
from several points of view: the brain and its structural features can
be seen as a prototype of a physical system capable of highly complex
and adaptable patterns in connectivity, selectively improved through
evolution; architectural organization of brain cortex is one of the
key features of how brain system evolves, adapts itself to the
experience, and to possible injuries.\\ Brain activity can indeed be
modelled as a dynamical process acting on a network; each vertex of
the structure represents an elementary component, such as brain areas,
groups of neurons or individual cells. A measure, called complexity,
has been introduced \cite{ton} with the purpose to get a sensible
measure of two important features of the brain activity: segregation
and integration.  The former is a measure of the relative statistical
independence of small subsets; the latter is the measure of
statistical deviation from independence of large subsets.\\ Complexity
is based on the values of the Shannon entropy calculated over the
dynamics of the different sized subgraphs of the whole network.  It is
sensitive both to the statistical properties of the dynamics and to
the connectivity.\\ It has been shown \cite{ton0}, by means of genetic
algorithms, that the graphs showing high values of complexity are
characterized by being both segregated and integrated; the complexity
is low when the system is either completely independent (segregated),
or completely dependent (integrated).  This general behaviour is valid
over a wide range of dynamical processes
\cite{ton0,ton1,ton2,ton3}. 


Despite this evidence, analytical results about the de\-pen\-den\-ce of
com\-ple\-xity on the to\-po\-logy and the dy\-na\-mics is still lacking.\\ 
In the
following will be proposed a first approach to this problem when the
dynamics is gaussian. 
The use of the gaussian dynamics get the statistical measure of 
complexity independent from the dynamics itself.
It doesn't pretend to represent any realistic
brain structure or activity, but to offer a first basic step for
understanding the relation existing between values of complexity and
topological properties of brain structure.
For this reason we used a simplified version of the model introduced by 
\cite{ton1}. This is a first step which could be furtherly developed 
for example for directed and weighted graphs.
\\

II.DYNAMICS, ENTROPY AND COMPLEXITY\\


We
consider a graph composed by $n$ vertices and $m$ links. It can be
represented by its adjacency matrix $\hat{\textbf{A}}$, whose elements
$a_{ij}$ we set to $1$ if there is a link between the vertices $i$ and
$j$, and $0$ otherwise.  Only non self connections are considered and
$a_{ij}=a_{ji}$.\\ 
On the graph we model the activity as a stochastic
process in the following way: each node $i$ at time $t$ 
can be in a particular state defined by the quantity 
$X_i(t)$. Given our graph with $n$ nodes, the states of the whole graph 
at time $t$ is given by the $n-dimensional$  vector $\textbf{X}(t)$. 

The evolution of $\textbf{X}(t)$ is given by the following dynamics:

\begin{eqnarray}
\label{dyn}
\mathbf{X}(t+1) &=& \hat{\mathbf{C}} \cdot
\mathbf{X}(t)+\mathbf{R}(t) \nonumber\\ \mathbf{X}(0) &=&
\mathbf{R}(0)
\end{eqnarray}

\bigskip

where $\hat{\textbf{C}}=\hat{\textbf{A}}/n$ and $\mathbf{R}(t)$ is an
$n-dimensional$ vector whose components $R_i(t)$ are random values.
$R_i(t)$ is chosen to be a white gaussian noise, i.e.  with the
following properties :

\begin{eqnarray}
\overline{R}_i(t) &=& 0\nonumber\\
\overline{R_i(t)R_j(t')} &=& \sigma \delta_{ij}\delta(t-t')
\end{eqnarray}

\bigskip
\bigskip

Here the bar represents the average over the ensemble.\\ The
normalization $n$ of the adjacency matrix assures that the process
will reach a stationary state.  The dynamics, described by
eq.(\ref{dyn}), is indeed a random walk which is damped if the matrix
has eingeinvalues $|{\lambda}|<1$.  In such a way the dynamics reaches
a stationary state with a caracteristic time $\tau \approx
1/\lambda_{min}$, where $\lambda_{min}$ is the smallest eingenvalues
of the matrix $\hat {\mathbf{C}}-\hat{\mathbf{I}}$.\\ 
It is
worth noting that the equation (\ref{dyn}) represents a simplified
version of the dynamics introduced in \cite{ton1}. In that case it was
considered a gaussian dynamics on directed and weighted graphs and with
some limitations on the values of the variances of each unit (node).


\bigskip

Since  $\textbf{X}(t)$ is a multidimensional gaussian process, its statistics 
is completely described through its second order moment: 

\begin{equation}
\label{dynamics}
\overline{\mathbf{X}(t+1)\mathbf{X}^t(t+1)}=\hat{\mathbf{C}}\overline{\mathbf{X}(t)\mathbf{X}^t(t)}\hat{\mathbf{C}}^t+\hat{\mathbf{I}}
\end{equation}

\bigskip

$\overline{\mathbf{X}(t+1)\mathbf{X}^t(t+1)}$ is the $n\times n$
covariance matrix whose determinant will be referred in the following
as $|cov(\mathbf{X})|$. The average value of $\textbf{X}(t)$
is always zero being a sum of zero mean values at each time step.\\
Since the process $\mathbf{X(t)}$ is gaussian, 
it is possible to show that the Shannon entropy $H(\mathbf{X})$ depends
only on $|cov(\mathbf{X})|$ \cite{pap}:

\begin{equation}
\label{entropy2}
H(\mathbf{X})=0.5 \cdot \ln [{(2\pi e)^n|cov(\mathbf{X})|}]
\end{equation}

\bigskip

Let us consider all the possible subgraphs of rank $\it{k}$ (number of
nodes) of the whole graph. Each of these subgraphs are indicated as
$\mathbf{X}_k$.
 
The complexity has been defined as:

\begin{equation}
\label{complexity}
C(\mathbf{X})=\sum_k \left[\langle
H(\mathbf{X}_k)\rangle-\frac{k}{n}H(\mathbf{X})\right]
\end{equation}

where the average $\langle ...\rangle$ is taken over all the subgraphs
of rank $k$. The sum ranges from the minimum possible rank of a
subgraph, i.e. $2$ to $n-1$. The term in (\ref{complexity}) for $k=1$
would be trivial since the covariance matrix of disconnected vertices
is simply dependent only on the variance of $\mathbf{R}(t)$,
i.e. $|\mathbf{COV(\mathbf{X}_1)}|=\sigma^n$; the term for $k=n$ is
instead always null. In the following, we will set for the sake of 
simplicity $\sigma = 1$.

In what follows we will try to find a relation between the to\-po\-lo\-gy of
the graph and its values of Entropy $H(\mathbf{X})$ and Complexity
$C(\mathbf{X})$, having defined on it the multidimensional gaussian
process (\ref{dyn}).\\

Under stationary conditions, the generic element of $cov(\mathbf{X})$
in the eigenvectors base {$\textbf{x}'_i$} is:

\begin{equation}
\label{cov}
\overline{x'_i x'_l}=\frac{\delta_{il}}{1-\-\frac{\lambda_i} {n}
\frac{\lambda_l} {n}}
\end{equation}

\bigskip
\bigskip

The set of value $\lambda_i$ represents the eigenvalue spectrum of the
adjacency matrix $\hat{\textbf{A}}$:

\begin{equation}
\label{eigenvalue}
\sum_j A_{ij}x'_j=\sum_j \lambda_j x'_j \delta_{ij}
\end{equation}

Following (\ref{cov}) the determinant of the covariance matrix is:

\begin{equation}
\label{det}
|cov(X)|=\prod_{i} \frac{1}{1-\frac{\lambda_i^2} {n^2}}\\
\end{equation}

This expression shows that the dynamics depends only on the properties
of the adjacency matrix $\hat{\textbf{A}}$ through its eigenvalue
spectrum. As a consequence the statistical properties of the
stationary states can be analized without studying their time evolution but
by looking at their eigenvalue spectrum.
On the other hand the richness in information embedded
in the eigenvalue spectrum makes the analysis not trivial at all
\cite{spectra}. The aim of the next paragraph is to show which
topological properties embedded into the spectrum dominate the
behaviour of the dynamical process.

\bigskip

III. CONNECTION WITH THE TOPOLOGY\\

Using the equation (\ref{det}), $H(\mathbf{X})$ becomes:
\bigskip

\begin{eqnarray}
\label{entropy3}
H(\mathbf{X}) &=& 0.5\cdot \ln[(2\pi e)^n|cov(\mathbf{X})|]
     \nonumber\\ &=& 0.5\cdot \sum_{i}\ln\left[(2\pi
     e)\frac{1}{1-\frac{\lambda_i^2} {n^2}}\right]
\end{eqnarray}

\bigskip

If $\lambda_{max}^2/n^2 << 1$, 
we can consider the following series expansion: 

\begin{eqnarray}
\label{log}
\ln\left(1-\frac{\lambda_i^2}{n^2}\right) & \simeq 
&-\sum_i\frac{\lambda_i^2}{n^2}-\sum_i\frac{\lambda_i^4}{2 n^4}+ \nonumber\\
&+& O\left(\sum_i \frac{\lambda_i^6}{3 n^6}\right)
\end{eqnarray}

\bigskip

and by substitution in (\ref{entropy3}), we get:

\begin{eqnarray}
\label{entropy_expan}
H(\mathbf{X})& \simeq & \ln(2 \pi e)^{\frac{n}{2}}+
0.5 \left(\sum_i\frac{\lambda_i^2}{n^2}+\sum_i\frac{\lambda_i^4}{2
n^4}\right) + \nonumber\\
&+& O\left(\sum_i\frac{\lambda_i^6}{6 n^6}\right)
\end{eqnarray}

\bigskip

Eq.(\ref{entropy_expan}) allows us to relate $H(\mathbf{X})$ to the number
$D_k$; this is the number of $k-step$ directed paths of the underlying
-undirected- graph, which return to their starting node after $k$
steps:

\begin{eqnarray}
\label{trace}
D_k=\sum_{i=1}^{n}(\lambda_i)^k=\sum_{i_1,i_2,...,i_k}a_{i_1,i_2}a_{i_2,i_3}...a_{i_k,i_1}
\end{eqnarray}

where $a_{i_k,i_{k+1}}$ is the generic non zero element of the
adiancency matrix $\hat{\mathbf{A}}$.

\bigskip
\bigskip

Using this result $H(\mathbf{X})$ becomes:

\begin{eqnarray}
\label{entropy_exp2}
H(\mathbf{X}) &\simeq& \ln(2 \pi
e)^{\frac{n}{2}}+0.5\left(\frac{D_2}{n^2}+\frac{D_4}{2n^4}\right)+\nonumber\\
&+&O\left(\frac{D_6}{6 n^6}\right)
\end{eqnarray}

\bigskip

$D_2$ is then the number of paths which starting from 
any node $i$ go to any other one $j$ and then come 
back to $i$. Remembering that an unconnected pair of 
nodes has $a_{ij}=0$, $D_2$ is obviously twice the number of {\it links} 
of the whole graph.

Thus the first two terms of $H(\mathbf{X})$ ex\-pan\-sion de\-pend on\-ly on
the num\-ber of nodes and links, and not on the graph topology.\\

Consider now the value of complexity $C(\mathbf{X})$ up to the $D_2$
term in the entropy.  We get

\begin{eqnarray}
\label{primotermine}
C(\mathbf{X}) &=& \sum_{k=2}^{n-1} \left[\langle
H(\mathbf{X}_k)\rangle -\frac{k}{n}H(\mathbf{X})\right] \simeq
\nonumber\\ & \simeq & 0.5 \sum_{k=2}^{n-1}\left(\frac{\langle
D_2(k)\rangle}{k^2}-\frac{k}{n}\frac{D_2}{n^2}\right)=\nonumber\\ &=&
\sum_{k=2}^{n-1}\left(\frac{\langle
m(k)\rangle}{k^2}-\frac{m}{n^2}\frac{k}{n}\right)=\nonumber\\ &=&
\frac{m(n-2)}{n(n-1)}-\frac{m(n+1)(n-2)}{2n^3}+\frac{m}{n^3}\sum_{k=2}^{n-1}\frac{1}{k}=\nonumber\\ &=& C^{ord2}(n,m)
\end{eqnarray}

\bigskip

since:

$$
\langle m(k)\rangle=m\frac{k(k-1)}{n(n-1)}
$$

\bigskip
\bigskip

So far, the value of $C(\mathbf{X})$ is not defined by the topology.
In order to reveal something related to a particular link's
arrangement, we need to consider the further terms in the expansion.
We can rewrite the complexity $C(\mathbf{X})$ in the following way to
put in evidence the part dependent only on the number of links $m$ and
nodes $n$ of the whole graph 
(and so independent from the topology), $C^{ord2}(n,m)$:

\begin{eqnarray}
\label{remembering}
C(\mathbf{X}) &=& C^{ord2}(n,m)+\nonumber\\ &+&\frac{1}{4} \sum_{k=2}^{n-1}
\left(\frac{\langle
D_4(k)\rangle}{k^4}-\frac{k}{n}\frac{D_4}{n^4}\right)+R(\lambda)
\end{eqnarray}

\begin{equation}
\label{resto}
R(\lambda)=O\left[\sum_{k=2}^{n-1} \left(\sum_i \frac{\langle
\lambda_i(k)^6\rangle}{6 k^6}\right) -\frac{k}{n} \frac{D_6}{6
n^6}\right]
\end{equation}

\bigskip

where we have explicitly written the term in $\lambda^4$.

\bigskip
\bigskip

In order to express the 
topological information contained in the eq.(\ref{remembering}), 
let us consider that


\begin{eqnarray}
\label{variance}
\langle D_4(k)\rangle
&=&\langle\sum_i \lambda_i(k)^4\rangle=\nonumber\\
&=&\langle 4 m(k)^2\rangle-\sum_{i\neq j}\langle\lambda_i(k)^2
\lambda_j(k)^2\rangle
\end{eqnarray}

and

\begin{equation} \label{sum}
\sum_{i \neq j} \lambda_i^2(k) \lambda_j^2(k)=\sum_{i \neq j} \sum_{lq}
a_{il}a_{li}a_{jq}a_{qj}-2D_2(k)
\end{equation}

\bigskip

From eq.(\ref{variance}) we see that the $C(\mathbf{X})$, at this or\-der
of ap\-pro\-xi\-ma\-tion, de\-pends on the se\-cond order 
moment of the number of
links $\langle m(k)^2 \rangle$, calculated over all the subgraphs of
rank $k$ (we remember that a subgraph of {\it rank $k$} is a particular 
choice of $k$ nodes in the whole graph and  $k \in [2, n-1]$).  
This is the first quan\-ti\-ty
de\-pen\-dent on the to\-po\-lo\-gy that we can easily e\-va\-luate 
on the graph in
place of the original expression of complexity.  In the next step we
will show that the fourth order approximated value of complexity can
be expressed through $\langle m(k)^2 \rangle$ calculated over all the
subgraphs of rank $k$, (i.e. $k \in [2, n-1]$) and the second order
moment of the degree distribution $\langle \langle q^2\rangle\rangle$ of the
whole graph.  This will allow to distinguish the complexity of graphs
with the same total number of nodes $n$ and total number of links 
$m$, through the evaluation of less time
consuming measures than the complexity. Moreover this result will
offer a deeper understanding of what the complexity measure means from the
topological point of view. \\

The terms in the sum of eq.(\ref{sum}) can be explicitly expressed as:

\begin{equation}
\label{sum_loop}
\begin{array}{lcl}
&&\sum_{i \neq j} 
\sum_{lq} a_{il}a_{li}a_{jq}a_{qj} =\\
&&\\
&=& \sum_{i \neq j; l=j,q=i}
(a_{ij}a_{ji})^2 + \sum_{i \neq j \neq l \neq q} 
a_{il}a_{li}a_{jq}a_{qj} +\\
&&\\
&+& 3 \sum_{i \neq j \neq l} a_{il}a_{li}a_{jl}a_{lj} + 
2  \sum_{i \neq j \neq l}a_{il}a_{il}a_{jj}a_{jj} +\\
&&\\
&+& \sum_{i \neq j; l=i, q=j} a_{ii}a_{ii}a_{jj}a_{jj} + 2 
\sum_{i \neq j l=q=j} a_{ij}a_{ji}a_{jj}a_{jj}\\
\end{array}
\end{equation}

\bigskip

In the expression (\ref{sum_loop}), only 
the first three terms are non zero, while the others
contain at least a diagonal element $a_{ii}=0$. Moreover the first
term is just $D_2$.

It is easy to show that the second and the third terms in the sum
correspond to the number of paths (``loops'') of the type shown in Fig.
\ref{loop}:

\begin{figure}[htbp]
\begin{center}
\includegraphics[height=5cm,width=6cm]{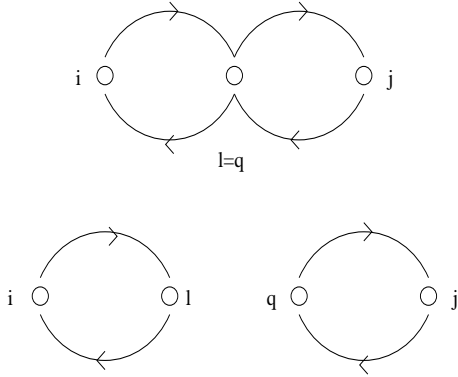}
\end{center}
\caption{{\it top}: paths described by the third term in eq.(\ref{sum_loop}); 
{\it bottom}: paths described by the second term in eq.(\ref{sum_loop})}
\label{loop}
\end{figure}

We will show that the number of these paths is
re\-la\-ted to $\langle \langle q^2\rangle \rangle$.  In
what follows $\it{q}$ represents the degree of the generic node
$\it{i}$, i.e. the node $\it{i}$ has $\it{q}$ links.\\

Consider now a particular subgraph of rank {\it k}. It has {\it k} 
nodes, each of them having a certain number of links or none.
Consider then all the nodes having the same degree $\it{q}$ in a
generic subgraph of rank $\it{k}$; then the number of paths (loops) of the
first type in Fig.\ref{loop}, involving this kind of nodes are:
 
\begin{equation}
{n-3 \choose k-3} \cdot {q \choose 2} \nonumber
\end{equation}

where ${n-3 \choose k-3}$ is the number of ways to choose $\it{k}$
nodes over a total of $\it{n}$ nodes, leaving aside the 3 nodes which
belong to the pair considered. 
This is the number of subgraph of rank {\it k}, in the whole graph, 
which contain a particular choice of 3 nodes.
Moreover, since the generic node
$\it{i}$ has $\it{q}$ links, we can count ${q \choose 2}$ different
pairs of links sharing the same node $\it{i}$. \\If we denote with
$P(q)$ the degree distribution in the whole graph, then we can write
the third term in eq.(\ref{sum_loop}) as :

\begin{eqnarray}
\sum_{\{k\}} \sum_{i \neq j \neq l}
a_{il}a_{li}a_{jl}a_{lj}&=&\int_{1}^{q_{max}}{n-3\choose k-3}{q\choose
2} P(q) dq=\nonumber\\ &=&\frac{1}{2}{n-3\choose k-3}(\langle \langle
q^2\rangle \rangle-\langle \langle q\rangle \rangle)\nonumber
\end{eqnarray}
where we explicitely wrote the $\sum_{\{k\}}$ which we will use later. 
 $\sum_{\{k\}}$ means the sum over all the subgraphs with $\it{k}$
nodes or of rank {\it k}.\\ Consider now the second term in 
eq.(\ref{sum_loop}), i.e. $\sum_{i \neq j \neq k \neq
q}a_{ik}a_{ki}a_{jq}a_{qj}$.  This is the number of disjoint pairs of
links in a generic subgraph of rank $\it{k}$. \\To compute such
number, we first count the total number of pairs in the whole graph,
i.e. ${m \choose 2}$; then we subtract the number of pairs of links
sharing a node in the whole graph (from the previous
computation). Finally we have to consider the multiplicity ${n-4
\choose k-4}$, i.e. the number of subgraphs of rank $\it{k}$
containing each pair of disjoint links.\\ Then the second term in
eq.(\ref{sum_loop}) is (again considering also the sum: $\sum_{\{k\}}$) :

\begin{eqnarray}
\sum_{\{k\}}\sum_{i \neq j \neq l \neq q}
a_{il}a_{li}a_{jq}a_{qj} &=& \left[{m \choose 2
}-\frac{1}{2}(\langle\langle q^2 \rangle\rangle-\langle\langle q
\rangle\rangle)\right]\nonumber\\
&\cdot& {n-4 \choose k-4} \nonumber
\end{eqnarray}

Remembering that in eq.(\ref{remembering}) we have to compute the
quantity $\sum_{k=2}^{n-1} \langle D_4(k)\rangle$, then we have to evaluate
$\sum_{k=2}^{n-1}\sum_{i \neq j} \sum_{lq} \langle
a_{il}a_{li}a_{jq}a_{qj}\rangle$. The average $\langle ...\rangle $ is 
performed for a particular value of {\it k} 
over all the subgraph of rank {\it k}.
For this reason in previous expressions 
we have explicitely written the sum over all 
the subgraphs of rank {\it k}, i.e. $\sum_{\{k\}}$.
To perform the average $\langle ...\rangle $ we have then simply to divide 
such expressions for the number of subgraph of rank {\it k} 
contained in the whole graph, i.e.$n \choose
k$.
Eventually we have to sum over all the values of {\it k},
$\sum_k$.\\

 Then the final expression for $\sum_k \langle D_4(k)\rangle$ becomes:

\begin{equation}
\begin{array}{lcl}
\sum_k \langle D_4(k)\rangle=\sum_k \langle 4 m(k)^2
\rangle+\sum_{k=2}^{k=n-1}\frac{\langle 2 m(k)\rangle}{{n \choose
k}}\\ &\\ +\frac{1}{2}\sum_{k=4}^{n-1}(\langle\langle
q^2\rangle\rangle-\langle\langle q\rangle\rangle) \frac{{n-4 \choose
k-4}}{{n \choose k}}-{m \choose 2}\sum_{k=4}^{n-1}\frac{{n-4 \choose
k-4}}{{n \choose k}}\\ &\\ +\frac{3}{2}\sum_{k=3}^{n-1}(\langle\langle
q\rangle\rangle-\langle\langle q^2\rangle\rangle)\frac{{n-3 \choose
k-3}}{{n \choose k}}
\end{array}
\end{equation}

\bigskip

where the second and third term of eq.(\ref{sum_loop}) have been
substituted with their explicit computations.\\ It is worth to note
that $\sum_{k=2}^{n-1} \langle\langle q \rangle\rangle$ is only a function of
$\it{n}$ and $\it{m}$, being $\langle\langle q \rangle\rangle=\frac{m}{n}$.

\bigskip

In the following is the whole expression of $C(X)$:

\begin{equation}
\label{final}
\begin{array}{lcl}
C(X)=C^{ord 2}(n,m)+C^{ord4}_1(m,n,\sum_{k=2}^{n-1}\langle
m(k)^2\rangle)+\\ 
&\\ + C^{ord4}_2(m,n,\langle\langle
q^2\rangle\rangle)+C^{ord4}_3(m,n, \langle\langle q \rangle\rangle)+\\ 
&\\
+ C^{ord4}_4(m,n)+R(\lambda)
\end{array}
\end{equation}

\bigskip
where $C^{ord 2}(n,m)$ and $R(\lambda)$ are respectively the
eq.(\ref{primotermine}) and eq.(\ref{resto}). The other terms are:

\begin{equation}
\begin{array}{lcl}
C^{ord4}_1(m,n,\sum_{k=2}^{n-1}\langle m(k)^2\rangle)=\sum_{k=2}^{n-1}
\frac{\langle m(k)^2 \rangle}{k^4}\\

&\\
&\\

C^{ord4}_2(m,n,\langle\langle q^2\rangle\rangle)=\frac{\langle\langle q^2
\rangle\rangle}{8}\frac{n^2-n-3}{n^4}+\\
&\\
+\frac{\langle\langle q^2 \rangle\rangle}{8}
\left(\sum_{k=4}^{n-1} \frac{1}{k^4}\frac{(n-4)!k!}{(k-4)! n!}
-\sum_{k=3}^{n-1}\frac{3}{k^4}\frac{(n-3)!k!}{(k-3)!n!}\right)\\

&\\
&\\

C^{ord4}_3(m,n,\langle\langle
q\rangle\rangle)=-\frac{1}{8}\frac{n^2-n-3}{n^4}+\\ &\\
+\frac{1}{8}\left(\sum_{k=3}^{n-1}\frac{3}{k^4}\frac{(n-3)!k!}{(k-3)!n!}-\sum_{k=4}^{n-1}
\frac{1}{k^4}\frac{(n-4)!k!}{(k-4)! n!} \right)\\

&\\
&\\

C^{ord4}_4(m,n)=-\frac{m(n+1)(n-2)}{4n^5}+\\ &\\
+\frac{m}{2}\left(\sum_{k=2}^{k=n-1}\frac{1}{n(n-1)}\frac{1}{{n
\choose k}}\frac{k-1} {k^3}-\frac{m-1}{4}\frac{{n-4 \choose k-4}}{{n
\choose k}}\frac{1}{k^4}\right)\nonumber\\
\end{array}
\end{equation}

\bigskip
\bigskip

\begin{figure}[htbp]
\begin{center}
\includegraphics[width=8cm]{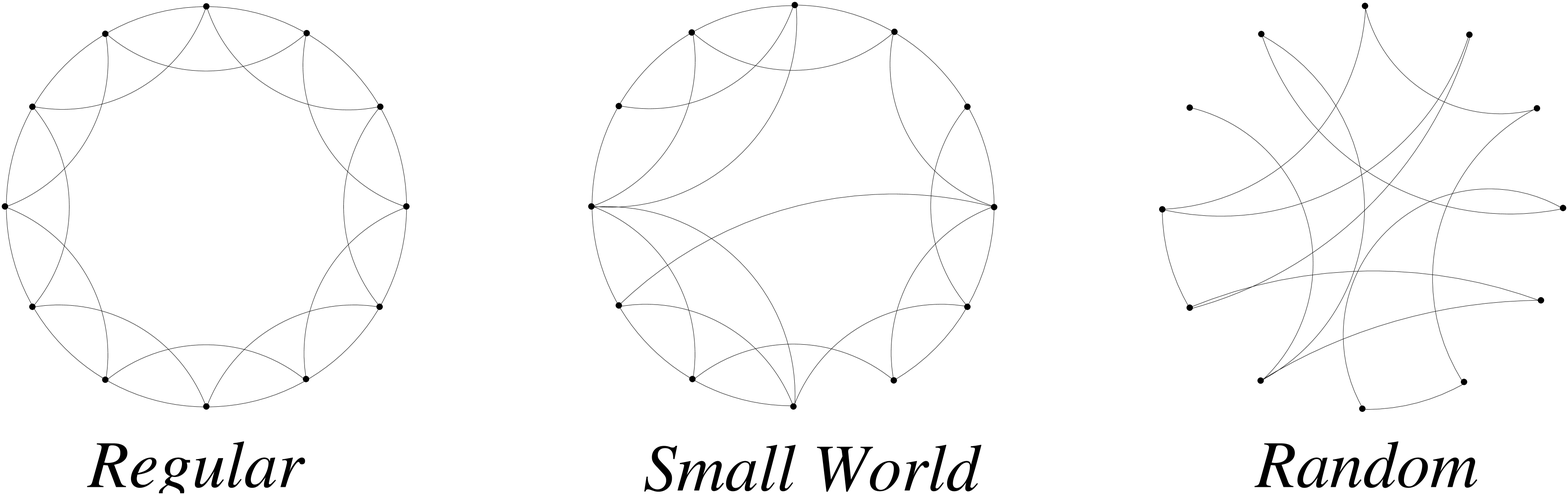}
\end{center}
\caption{3 different network models: a regular network, the small word model and a random network. It is possible to go from one model to the other varying the probability $p$ of rewiring (see the text for further details).\\}
\label{models}
\end{figure}

\bigskip
\bigskip

\begin{figure}
\begin{center}
\includegraphics[height=6.5cm,width=8cm]{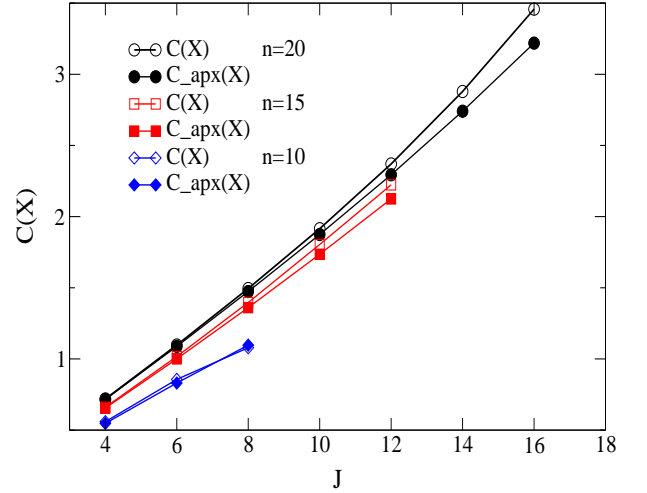}
\end{center}
\caption{Behaviour of complexity measure and its approximation (up to 
the fourth order) in a small world graph with n=10,15,20 nodes and p=0.1,
against J, i.e. number of first neighbors.\\}
\label{small}
\end{figure}
\begin{figure}
\begin{center}[htbp]
\includegraphics[height=6.5cm,width=8cm]{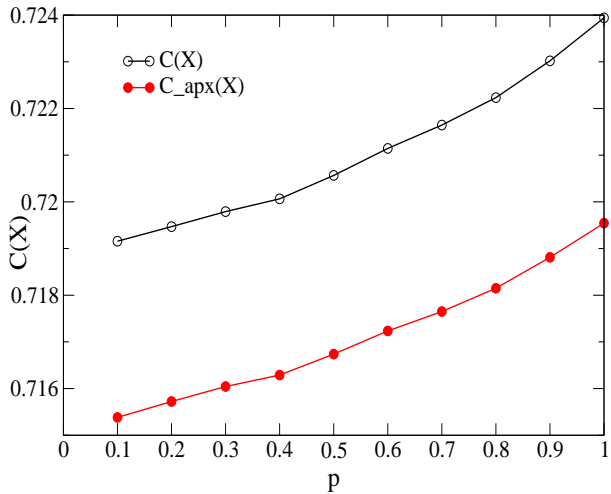}
\end{center}
\caption{Behaviour of complexity measure and its approximation (up to 
the fourth order) in a small world graph with n=20 nodes and J=4, against p,
probability of rewiring.\\}
\label{approx_exact}
\end{figure}

\bigskip
\bigskip

IV. NUMERICAL RESULTS\\

We perform calculation of entropy and complexity of the dynamics
(\ref{dyn}) over a small world graph with $n=10,15,20$ nodes. 
We estimate both the
exact and the approximate values, for checking the accuracy of the
approximation, and their dependence on the topological properties of
the graphs.

The algorithm behind the model can be summarized in two steps
\cite{wat}:\\ (1) Start with a ring lattice with $\it{n}$ nodes in
which every node is connected to its first $\it{J}$ neighbors
($\it{J/2}$ on either side). In order to have a sparse but connected
network at all times, consider $n>>J>>ln(n)>>1$.\\ (2) Randomly rewire
each edge of the lattice with probability $\it{p}$ such that self-connections
and duplicate edges are excluded. Varying $\it{p}$ the transiton between
order $(p=0)$ and randomness $(p=1)$ can be closely monitored 
(Fig.\ref{models}).

The numerical evaluation for the exact and the approximated values of
$C(\mathbf{X})$ can be easily achieved in a small world graph: in this
case we can investigate different arrangements of links keeping the
number of nodes and links fixed. The variation of complexity is
affected both by $\langle m(k)^2 \rangle$ over all the scales
$\it{k}$, and $\langle\langle q^2\rangle\rangle$.

Since the expansion is allowed when the average node degree is much
less than one ($\lambda_{max}^2/n^2 << 1$), we expect a higher
accuracy when the average connectivity is low ($n>>J$), and a worse
approximation when $\it{J}$ increases. The simulations confirm this
trend for increasing values of $\it{J}$, and $J<n$ (Fig. \ref{small}).\\ In
Fig.\ref{approx_exact} 
we show the exact and approximated behaviour of $C(\mathbf{X})$
versus the probability of rewiring.Their relative difference is much
less than one and they show very similar behaviour.\\ Analogous
results have been found for the other values of $\it{n}$.\\

V. DISCUSSION\\

We attempted to extract the topological meaning of the complexity
measure in the case of gaussian dynamic. This aim has been achieved
both from analytical and numerical points of view, showing that very
good approximation of complexity can be obtained through two simple
direct topological measures on the graph, namely, the second order
moment of the number of links $\langle m^2(k)\rangle$ over all the
scales $\it{k}$, and the second order moment of the node degree
$\langle\langle q^2\rangle\rangle$ on the whole graph.
The analytical expression is obtained through an expansion for 
$\lambda_{max}/n <<1$;
however the numerical results show the 
expression (\ref{final}) 
for the complexity is a reasonable approximation even for 
$\lambda_{max}/n \lessapprox 1$.\\
The relevance of the
obtained results relies on two main aspects: the measure has a clear
topological meaning which help to understand in a more intutive way
the degree of complexity of a graph; it can be evaluated through two
less time-consuming, and considerably easier to compute topological
measures. The saving of computation time is of order $\it{n}$ since
evaluating the two mesures mentioned above require $\it{n^2}$ steps
instead of the $\it{n^3}$ steps of the diagonalizing algorithms for
symmetric matrices.\\

We enjoyed useful discussions and suggestions by Dr.V.Servedio and Dr.A. Capocci.
We would like to thank Dr. F.Colaiori for her help in drawing figures.



\end{document}